%% file: main.tex
\documentclass[letterpaper,twocolumn,10pt]{article}
\usepackage{usenix-2020-09}

\usepackage{tikz}
\usepackage{amsmath}

\usepackage{filecontents}

\usepackage{booktabs}
\usepackage{multirow}
\usepackage{tikz}
\usepackage{xcolor}
\usepackage{xspace}
\usepackage{tikz}
\usepackage{pifont}
\usepackage[english]{babel}
\usepackage{blindtext}
\usepackage{lipsum}
\usepackage{colortbl} 
\usepackage[ruled,vlined,linesnumbered]{algorithm2e}
\usepackage{booktabs}

\usepackage{filecontents}
\usepackage{xspace}
\usepackage{xcolor}
\usepackage{hhline}
\usepackage{multirow}
\usepackage{soul}

\usepackage{array}
\usepackage{comment}
\usepackage{listings}
\usepackage{dsfont}
\usepackage{algorithmic}
\usepackage{xurl}
\usepackage{setspace}
\usepackage{graphicx}
\usepackage{caption}
\usepackage{ltablex}
\usepackage{subcaption}
\usepackage{ifluatex}

\usepackage{outlines}

\newtheorem{insight}{{\bf Insight}}

\input{macros}

\title{\name: KV Cache Sharing for Cross-LLM Communication \\and Multi-LLM Serving}

\author{
  \normalsize
Yuhan Liu$^1$
\hspace{0.3em}
Yuyang Huang$^1$
Jiayi Yao$^1$
Shaoting Feng$^1$
Zhuohan Gu$^1$
Kuntai Du$^1$
Hanchen Li$^1$
Yihua Cheng$^1$\\
Junchen Jiang$^1$
\normalsize
Shan Lu$^2$
Madan Musuvathi$^2$
Esha Choukse$^2$
\\
\normalsize
$^1$University of Chicago\hspace{.4in}$^2$Microsoft
}
\begin{document}

\renewcommand{\title}[1]{}
\maketitle

\pagestyle{plain}


\input{sec-abstract}

\input{sec-intro5}

\input{sec-motivation4}
\input{sec-analyzeKV2}


\input{sec-design3}

\input{sec-eval2}

\input{sec-related2}
\input{sec-conclusion}

\bibliographystyle{plain}
\bibliography{citations}
\newpage
\appendix
\input{sec-appendix1}

\end{document}

%% file: macros.tex
\definecolor{darkkhaki}{rgb}{0.74, 0.72, 0.42}

\definecolor{yccolor}{RGB}{240, 24, 240}

\newcommand{\name}{DroidSpeak\xspace}

\newcounter{packednmbr}
\newenvironment{packedenumerate}{\begin{list}{\thepackednmbr.}{\usecounter{packednmbr}\setlength{\itemsep}{0.5pt}\addtolength{\labelwidth}{-4pt}\setlength{\leftmargin}{2ex}\setlength{\listparindent}{\parindent}\setlength{\parsep}{1pt}\setlength{\topsep}{0pt}}}{\end{list}}
\newenvironment{packeditemize}{\begin{list}{$\bullet$}{\setlength{\itemsep}{0.5pt}\addtolength{\labelwidth}{-4pt}\setlength{\leftmargin}{2ex}\setlength{\listparindent}{\parindent}\setlength{\parsep}{1pt}\setlength{\topsep}{2pt}}}{\end{list}}

\newcommand{\tightcaption}[1]{\vspace{-0.15cm}\caption{{\normalfont{\textit{{#1}}}}}\vspace{-0.3cm}}
\newcommand{\tightsection}[1]{\vspace{-0.1cm}\section{#1}\vspace{-0.00cm}}

\newcommand{\tightsubsection}[1]{\vspace{-0.3cm}\subsection{#1}\vspace{-0.02cm}}

\newcommand{\eg}{{\it e.g.,}\xspace}
\newcommand{\ie}{{\it i.e.,}\xspace}
\newcommand{\etal}{{\it et.~al}\xspace}

\newcommand{\mypara}[1]{\vspace{0.05cm}\noindent{\bf {#1}:}~}

\definecolor{backcolour}{rgb}{0.96,0.96,0.96}
\definecolor{codegray}{rgb}{0.5,0.5,0.5}
\definecolor{deepblue}{rgb}{0,0,0.6}
\definecolor{deepred}{rgb}{0.6,0,0}
\definecolor{deepgreen}{rgb}{0,0.5,0}
\lstdefinestyle{mystyle}{
    backgroundcolor=\color{backcolour},   
    commentstyle=\color{codegreen},
    morekeywords={self, True},
    keywordstyle=\color{deepblue},
    numberstyle=\tiny\color{codegray},
    emph={MyClass,__init__,EncodingType,Image},
    emphstyle=\color{deepred},
    stringstyle=\color{deepgreen},
    basicstyle=\ttfamily\footnotesize,
    breakatwhitespace=false,         
    breaklines=true,                 
    captionpos=b,                    
    keepspaces=true,                 
    numbers=left,                    
    numbersep=5pt,                  
    showspaces=false,                
    showstringspaces=false,
    showtabs=false,                  
    tabsize=1
}

\newcommand{\vminfive}{\vspace{-5pt}}
\newcommand{\vminten}{\vspace{-10pt}}

%% file: sec-abstract.tex
\begin{abstract}
{\em Compound AI systems}, such as agentic systems, 
are an emerging trend in large-scale enterprise settings, with multiple LLMs specialized for different users, tasks, and/or roles working together. 
In these scenarios, different models often process inputs that share the same context prefix. 
Although much work was done in the past to enable the reuse of prefix KV caches across inputs for a single model,
how to enable one model to reuse the prefix KV caches of a different model remains an open question. 


We introduce \name, the first distributed LLM inference system that enables KV cache reuse across distributed nodes running inference of different LLMs, so long as the LLMs have the same architecture.
We present the first study that aims at understanding the impact of sharing KV caches across different LLMs, and if/when such sharing affects quality. 
Inspired by the findings, we present \name, which selectively recomputes a few layers of the KV cache produced by another LLM and reuses the remaining
layers, with negligible quality loss. Moreover, carefully pipelining the layer-wise re-computation and the loading of reused KV cache further improves the inference performance.
Experiments on diverse datasets and model pairs demonstrate that \name achieves up to 4$\times$ throughput improvement and about 3.1$\times$ faster prefill (time to first token), with negligible loss of quality in F1 scores, Rouge-L or code similarity score, compared to the baseline which does not allow any sharing across models. 


\end{abstract}

%% file: sec-intro5.tex
\tightsection{Introduction}

Nowadays, LLM inference has become one of the most resource-consuming workloads in industry, demanding ever larger clusters of GPU machines~\cite{vllm_production_stack_2025, dynamo_2025, aibrix_2025, wu2024fastdistributedinferenceserving, patel2024splitwiseefficientgenerativellm, zhong2024distservedisaggregatingprefilldecoding}.
To reduce the computation demand, a common optimization is for GPU machines that run the {\em same} LLM to share KV caches of reused input prefixes over the network~\cite{chen2024kvdirectdistributeddisaggregatedllm, hu2024memservecontextcachingdisaggregated, li2024llminferenceservingsurvey, liu2024cachegenkvcachecompression}.
However, how to make that optimization work across {\em different} LLMs is yet to be studied.

Indeed, an emerging trend is hosting multiple {\em different} LLMs in one GPU cluster.
The reason is that with LLMs used in more complex or personalized tasks, multiple LLMs, often fine-tuned from the same foundational model, are needed to serve different users, tasks, or roles. These LLMs work together to perform one complex task or to offer different customized services~\cite{fang2024multiagentfinetuningapp, mineiro2024finetuningapp, borzunov2023distributedinferencefinetuninglarge, yang2024enterprise, arslan2024enterprise}.
Since standard prefix-caching techniques work only when the KV cache is reused by the same LLM, the use of multiple LLMs poses a direct challenge --- how to reuse KV cache efficiently across different LLMs in distributed settings. 

To motivate this challenge, we highlight three common use cases of multiple fine-tuned models working together in a system. 
The first one is \emph{multi-agent systems}, which use different fine-tuned models to serve as the agents and to accomplish collaborative tasks~\cite{chen2023fireactlanguageagentfinetuning, hong2024metagpt}. 
The second use case is serving \emph{multi-LoRA or multiple fine-tuned models}, such as models that are continuously updated over time with newer data \cite{huang2025alchemistdesignefficientonline}, or concurrently serving LoRA adapters~\cite{chen2023punicamultitenantloraserving, xia2024efficient, sheng2024sloraservingthousandsconcurrent}. 
The third use case is \emph{personalized assistant systems}, where a model is fine-tuned for each user (or user type) according to their personal preferences in coding or writing.  



\begin{figure}[t]
\centering
    \includegraphics[width=0.98\linewidth]{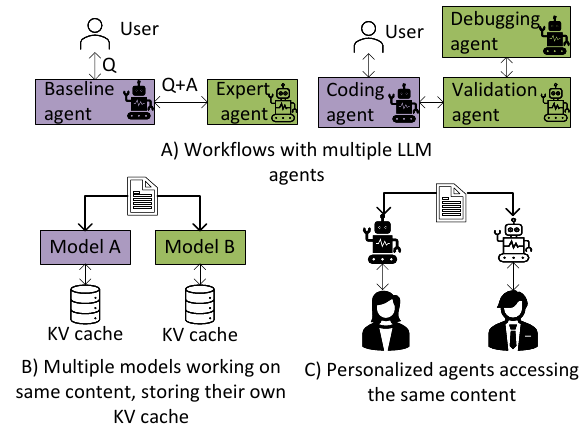}
    \tightcaption{
   Various scenarios in which same context is shared by multiple LLMs. \name brings down the computation latency by up to 3.1$\times$, increases throughput by 4$\times$.  
}
    \label{fig:overall}
\end{figure}

In all of these scenarios, prefix sharing is common. 
For example, in multi-agent systems (Figure~\ref{fig:overall}A), when the coding agent talks to the validation agent, the conversation history of the coding agent will be prepended to the input of the validation agent, to ensure the coherence of the conversation. 
In a multi-LLM or multi-LoRA inference system (Figure~\ref{fig:overall}B), an updated model in a chatbot application could refer to the same piece of conversation history produced by an older version of the model. 
In a personalized assistant system (Figure~\ref{fig:overall}C), where each assistant is fine-tuned for a different user's preference, the same news (\ie the query prefix) can be used to answer different users' queries. 



Based on the 
observations above, we pose the question: \emph{can we share the intermediate states (i.e., KV cache) produced by one LLM on a given context to accelerate the prefill for another LLM?} 
In this paper, we seek to answer this question under the assumption that the two LLMs have different weights but the same architecture.

Intuitively, the potential performance benefit of such intermediate state sharing is huge --- previous work on
single-LLM KV cache sharing has shown up to 8$\times$ latency and throughput improvement by speeding up the expensive prefill phase of LLM inference, particularly for long context workloads \cite{gao2024costefficientlargelanguagemodel, gim2024promptcachemodularattention, liu2024cachegenkvcachecompression, jin2024ragcacheefficientknowledgecaching}. 
However, just like a video decoder could not decode a video encoded with another codec, naively sharing the KV cache across different LLMs would cause generation quality to drop greatly.

Our hypothesis is that there should be a way to {\em re-compute a small portion and reuse a large portion}, although not all, of the KV cache 
between two models that are {\em fine-tuned from the same base model} --- since the models are only {\em fine}-tuned, they should share similar understanding of the same input context and hence help accelerate each other's
inference. Of course, the challenge is to validate this hypothesis, and to figure out which part to re-compute or
re-use without incurring much delay overhead or degradation of quality.

Through a thorough empirical study (\S\ref{subsec:empirical}), we measured eight representative model pairs and found that only a small subset, often around 10\%, of layers are sensitive to the KV cache difference between two models in a pair.
The identities of these layers vary with different model pairs, but are largely consistent across
different inputs for the same model pair. 
We refer to these layers as {\em critical layers} in this paper. 
Therefore, for each pair of LLMs, we propose to selectively recompute these critical layers in the KV cache. 

We build our insights into {\em \name}, the first distributed multiple-LLM inference system that enables efficient sharing of KV caches across different LLMs.
First, \name identifies the critical layer groups through offline profiling on a held-out ``training'' set, ensuring sufficient re-computation for accuracy purposes while reusing as many layers' KV cache as possible.
Second, \name implements smart KV cache loading, which pipelines the loading of KV cache with the re-computation of critical layers, to hide the loading delay from remote nodes as much as possible. 

We should note that the high-level idea of selectively recomputing KV cache is not exactly new. \name differs with prior work \cite{gao2024costefficientlargelanguagemodel, gim2024promptcachemodularattention, liu2024cachegenkvcachecompression, yao2024cacheblendfastlargelanguage} in
{\em why} and {\em how} to re-compute KV cache: \name is designed for KV cache reuse across
{\em different} LLMs, while prior work assumes a {\em single} LLM; due to the different purposes, 
none of the prior work chooses to re-compute or re-use a group of layers like in \name. 
For example, CacheBlend~\cite{yao2024cacheblendfastlargelanguage} updates a reused KV cache, but it still feeds the KV cache to the {\em same} LLM that generated the KV cache. Moreover, it updates the KV cache of certain tokens, rather than certain layers like in \name. 

We evaluate \name on six datasets across \emph{three} different tasks, including \emph{question answering, text summarization, and coding}, across eight different model pairs. 
We compare \name with various baselines, including direct KV reuse~\cite{gao2024costefficientlargelanguagemodel},  CacheBlend~\cite{yao2024cacheblendfastlargelanguage}, and smaller models. 
Across these setups, we can reduce the latency by 
up to 3.1$\times$, improve throughput by up to 4$\times$ with negligible drop in quality (measured in F1 score, Rouge-L, or code similarity score). 

While the concept of ``translating'' KV caches between models involves a machine-learning challenge, our primary contribution lies in making it practical in a {\em distributed system} setting, which requires the transfer and computation of such KV-cache translation to be done efficiently. 

We have integrated our implementation with LMCache~\cite{lmcache} and vLLM~\cite{kwon2023efficientmemorymanagementlarge}, each being the state-of-the-art KV cache management library and LLM inference engine, and it has been tested in enterprise settings. 
The link to the code is anonymized here for double-blind review requirements.




%% file: sec-motivation4.tex
\tightsection{Background \& Motivation}

In this section, we give a brief introduction to the background of the emerging workload of context sharing between different fine-tuned model versions and the motivation for \name.

\tightsubsection{Basic Transformer Concepts}
\label{sec:transformer}
The recent wave of generative AI is fueled by the advent of high-performing models that are transformer-based and decoder-only~\cite{hagos2024recentadvancesgenerativeai,ferrando2024primerinnerworkingstransformerbased,du2024valltdecoderonlygenerativetransducer}. 
\begin{figure}[t]
\centering
    \includegraphics[width=0.6\linewidth]{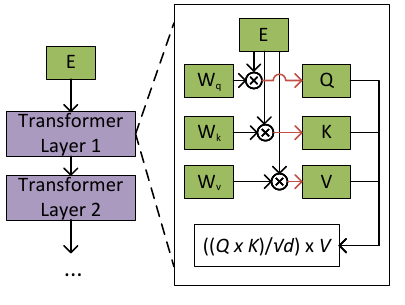}
    \vminten
    \tightcaption{
    Illustration of the use of embedding (E), query (Q), key (K), and value (V) tensors in self-attention in transformer-based LLMs.  
}
    \label{fig:kvcache-illustrated}
\end{figure}

\mypara{Query, Key, Value, and Embedding}
In transformers, Q (Query), K (Key), and V (Value) are the core components of the attention mechanism~\cite{vaswani2023attentionneed,niu2021review,brauwers2021general,lin2022survey,yeh2023attentionvizglobalviewtransformer}.
An LLM model comprises many layers. Each layer generates its E/Q/K/V tensors given an input (Figure~\ref{fig:kvcache-illustrated}). 
We denote the K and V tensors altogether as \emph{\textbf{KV cache}}, and the embedding E tensors as \emph{\textbf{E cache}}. 
Within each layer, embeddings E are the starting point for subsequent transformer computations (including attention). 

\textit{The quality of E/Q/K/V tensors directly affects the model's ability to understand and process the input context effectively.} 

\begin{figure*}

\begin{minipage}[b]{0.35\textwidth}
\centering
    \includegraphics[width=1\textwidth]{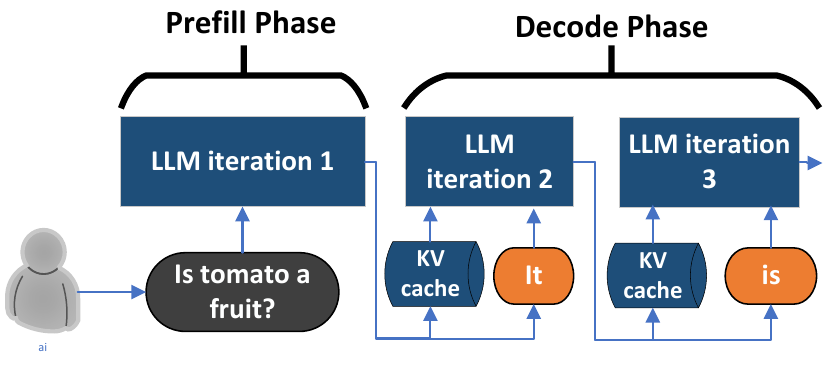}
   
    \tightcaption{Prefill and decode phases. }
    \vspace{10pt}
    \label{fig:prefill_Decode}
\end{minipage}
\hfill
\begin{minipage}[b]{0.26\textwidth}
\centering
    \includegraphics[width=0.9\textwidth]{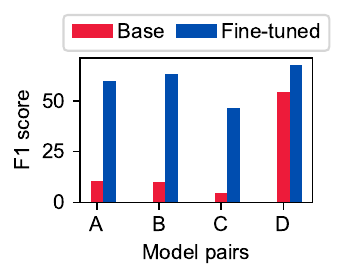}
    \vspace{-10pt}
    \tightcaption{ Fine-tuned model gives higher accuracy than baseline.
}
    \label{fig:finetuneperf}
\end{minipage}
\hfill
\begin{minipage}[b]{0.28\textwidth}
\centering
    \includegraphics[width=0.9\textwidth]{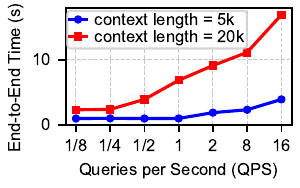}
    \vspace{-5pt}
    \tightcaption{ Shorter input leads to smaller end-to-end time.}
    \label{fig:tpt_motivation}
\end{minipage}

\end{figure*}
\mypara{Prefill and Decode phases}
\label{sec:prefilldecode}
LLMs process input and generate output in two distinct phases: the prefill phase and the decode phase, as shown in Figure~\ref{fig:prefill_Decode}.
In the \textbf{\textit{Prefill Phase}}, the LLM processes the entire input context to produce the embeddings and the KV caches across all layers. 
In the \textbf{\textit{Decode Phase}}, the model uses the KV cache generated in the prefill phase to autoregressively produce tokens one by one as the output. 

\mypara{Fine-tuned LLMs}
Despite being versatile, foundational LLMs' capabilities on specific tasks can improve through fine-tuning on specialized domain data. For example, one can
turn a foundational LLM into a customer-support agent by finetuning on troubleshooting requests \cite{Predibase}, or into a legal assistant by finetuning on
case law and statutes~\cite{yue2023disclawllmfinetuninglargelanguage}. Fine-tuning can greatly improve the accuracy of an LLM on a target domain, as shown in 
Figure~\ref{fig:finetuneperf}, where the fine-tuned models (\texttt{Llama-3-70B-Instruct~\cite{llama3modelcard}, Mistrallite~\cite{MistralLite_2023}, Llama-3-8B-Instruct~\cite{llama3modelcard}, and MAmmoTH2~\cite{yue2024mammoth2scalinginstructionsweb}}) greatly outperform the foundational models they originate from, on the HotpotQA dataset~\cite{hotpotqa}.

Recent works in parameter efficient fine-tuning,
like LoRA~\cite{hu2021loralowrankadaptationlarge} have 
made fine-tuned models even more accessible by updating part of the model weights, reducing computational and memory resources during fine-tuning.


\tightsubsection{Context Sharing Across LLMs}
\label{sec:trends}

In compound AI systems, prefix contexts\footnote{Since a shared context is often the prefix of different inputs, we use {\em prefix} (caching/sharing) and {\em context} (caching/sharing) interchangeably.} are often shared across different LLMs—either to enhance the coherence of the chat experience or to reference the same set of background documents.
In this paper, for convenience, we use the following terminology:
\begin{packeditemize}
\item {\bf \em Sender} model produces the KV cache of a context;
\item {\bf \em Receiver} model reuses the KV cache (with limited recomputation) of the reused context.
\end{packeditemize}

We describe several concrete use cases of context sharing between different LLMs below.

\mypara{Agentic Workflows}
Agentic workflows represent a paradigm shift in automation and collaboration in the LLM space~\cite{li2024agentsneed, hu2024automateddesignagenticsystems, xi2023risepotentiallargelanguage, li2024personalllmagentsinsights, manus2025, cursor2025, githubcopilot}. 
These workflows integrate multiple specialized LLM agents, each fine-tuned for specific tasks, to collaborate and solve complex, multi-step problems, or let them play different roles in agent debating. 
For example, prior works propose using fine-tuned models as different agents to improve the generation quality and generalization of agents~\cite{chen2023fireactlanguageagentfinetuning, song2024agentbank, chen2025atlasagenttuninglearning, chen2024agentflandesigningdatamethods}.  
Compared with using different prompts on the same model for different agents, using fine-tuned models as different agents can better improve output quality~\cite{chen2023fireactlanguageagentfinetuning, song2024agentbank}.  


\textbf{\textit{Need for context sharing: }}
In agentic workflows, different agents often share a common context, often the conversation history of other agents, to ensure the coherence and consistency between agents~\cite{hong2024metagpt, wu2023autogen, guo2024largelanguagemodelbased}. 
As a concrete example, in coding agentic workflows with a coding and a testing agent, the testing agent (receiver model) has to read both the input instructions and generated code from the coding agent (sender model) to write appropriate unit tests to meet user's needs ~\cite{hong2024metagpt, qian2024chatdevcommunicativeagentssoftware, wu2023autogen}. 



\mypara{Personalized Models}
Personalized models tailored to individual users or tasks are increasingly prevalent in LLM systems, particularly in applications like chatbots, virtual assistants, and recommendation engines~\cite{chen2024large,bhuiyan2024role,valavanidis2023artificial}. 
In these applications, different assistants are typically LLMs fine-tuned for different users' preferences~\cite{li2024personalllmagentsinsights, hong2024cogagentvisuallanguagemodel}.
For example, the personal assistant for a software engineer can be fine-tuned to generate high-quality and concise code snippets, while the personal assistant for a financial analyst identifies marketing angles in the documents without technical detail.

\textbf{\textit{Need for context sharing: }}
These models often share overlapping contexts, such as common conversation histories or shared knowledge bases, to ensure continuity and relevance. For instance, two assistants answering similar queries about current events will process the same top news. 

\mypara{Multiple-LLM or multi-LoRA serving} In chatbot applications, LLMs often require continuous updates to incorporate new information to provide up-to-date support and higher-quality answers for users~\cite{dam2024completesurveyllmbasedai, arcadinho2024automatedtestgenerationevaluate}. 
As an example, ChatGPT APIs release new API versions based on the same foundational model about every two months, which fine-tunes on the emerging new data~\cite{azureopenai2025}.
Furthermore, multiple LoRA adapters often need to be concurrently served~\cite{sheng2024sloraservingthousandsconcurrent, chen2023punicamultitenantloraserving} to accomplish different tasks or serve different users.

\textbf{\textit{Need for context sharing: }} In this case, the updated model (receiver model) often needs to re-process the same sets of popular contexts processed by the older model (sender model) before. 
Multiple LoRA adapters can share their KV cache when processing the same context. 

We motivate \name with these emerging trends in the workloads today that fuel the need for efficient context sharing across fine-tuned LLMs.

\tightsubsection{Distributed LLM Inference Systems}
As the demand for LLM inference continues to grow, it has become common to serve LLMs in a cluster of GPU nodes. 
Many companies have developed their own distributed inference systems, such as vLLM Production Stack~\cite{vllm_production_stack_2025},
NVIDIA Dynamo~\cite{dynamo_2025}, and ByteDance AIBrix~\cite{aibrix_2025}. 
Among all these frameworks, KV cache sharing across nodes is one of the most important features for reducing prefill computation and increasing overall throughput.
Specifically, when there are multiple requests to the \emph{same model} querying a common prefix context, these systems can transfer KV cache generated by another user request, either from another GPU node or through a centralized storage backend~\cite{jin2025computeloadkvcache, gao2024costefficientlargelanguagemodel, chen2024kvdirectdistributeddisaggregatedllm}.


However, current distributed inference systems have not optimized for multiple-LLM inference yet--they do not explore the opportunity to share KV cache across GPU nodes when the requests are querying \emph{different models}.

\tightsubsection{Prefill interference}

Given the distinct characteristics of the prefill and decode phases described in Section~\ref{sec:prefilldecode}, lengthy prefill phases can significantly reduce an inference system’s goodput—defined as the number of queries processed per second within a latency SLO~\cite{zhong2024distservedisaggregatingprefilldecoding}. This reduction occurs because TTFT (time to first token) grows super-linearly with input length, and the decoding phase cannot start until the prefill phase completes. Consequently, long inputs often turn prefill delays into the end-to-end bottleneck. For example, as shown in Figure~\ref{fig:tpt_motivation}—where Llama-3-8B runs on a single A100 GPU with synthetic input lengths of 5K and 20K tokens—setting a 3-second latency SLO reveals that increasing the input size by only 4× can reduce goodput by as much as 32×.

%% file: sec-analyzeKV2.tex
\begin{table}[h]
  \small 
  \centering
  \begin{tabular}{@{}ll@{}}
    \toprule
    Receiver Model             & Sender  Model            \\ \midrule
    evolcode-3.1-8b                    & toolace-3.1-8b                   \\
    glue\_sst2                         & conllpp               \\
    mistral-blitz & mistral-24b \\
    phi3.5-mini-instr-adapter-v2  & phi-3.5-mini-instr-task15 \\
    llama-3-8b-sft-lora-ultrachat & fingpt-llama-3-8b         \\
    llama-3-70b-instruct         & llama-3-70b        \\
    mistrallite                   & mistral-7b                \\
    llama-3.1-70b-instruct        & llama-3.1-70b             \\
    \bottomrule
  \end{tabular}
  \vspace{5pt}
  \tightcaption{The model pairs used in our paper. 
  For each pair of models, we use the datasets that meet the requirements listed in \S\ref{subsec:bench} (\ie the receiver model has better quality than the sender model). 
  }
  \label{tab:models}
\end{table}

\tightsection{ Reusing KV cache across LLMs}
\label{sec:reuse_kv}

A simple way to eliminate the overhead of repetitively re-computing the KV cache that another LLM has generated before is 
to reuse the KV cache produced by another model. As a concrete example, on an A100 GPU and using \texttt{Llama-3.1-8B‑Instruct}, reusing the KV cache for a 40K‑token input can reduce prefill latency from 4s to 0.08 seconds.

This naturally leads to the key question: What effect does directly reusing another LLM's KV cache have on generation quality?

In this section, we present the first empirical study of how reusing another model’s KV cache impacts output quality, and we further examine whether these effects vary across individual layers of the cache.

\begin{figure*}[ht!]
  \centering
  \includegraphics[width=0.75\linewidth]{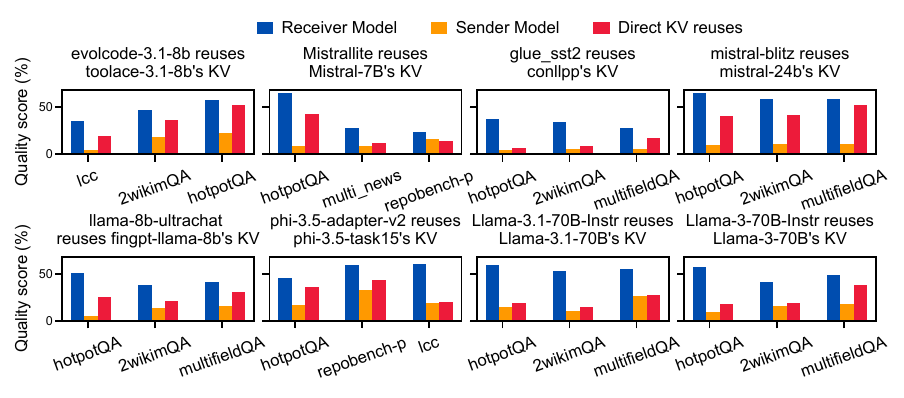}
  \vminten
  \tightcaption{
    Directly reusing the full KV cache greatly degrades generation quality.
    }
  \label{fig:directreuse}
\end{figure*}

\begin{figure*}[ht!]
  \centering
  \includegraphics[width=0.85\linewidth]{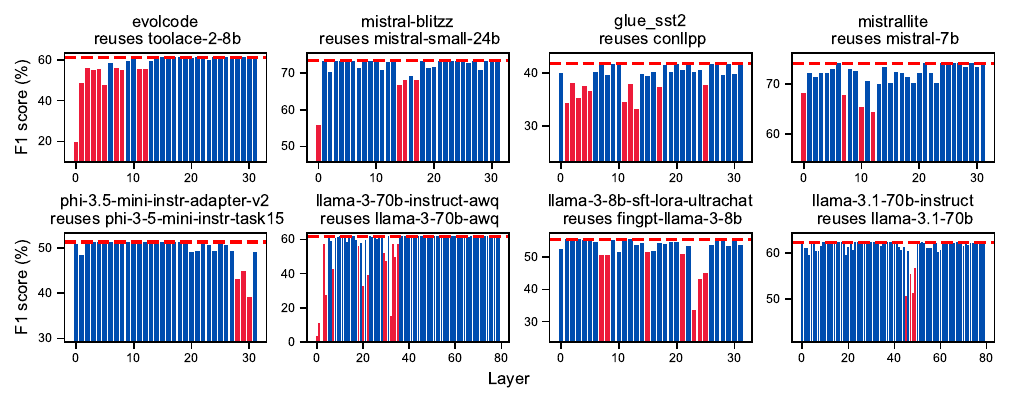}
  \vminten
  \tightcaption{
    Different layers have different sensitivities to deviation in KV cache.
    Plotted by reusing only one layer's KV cache from the sender model on the
    receiver model. The red dashed line is the original accuracy of the
    receiver model. The bars colored red are those that have an F1 score drop
    of over 10\% compared to the original receiver model. } \label{fig:layerwise_pattern_across_models}
\end{figure*}

\tightsubsection{Building the benchmarks and datasets }
\label{subsec:bench}
Before getting into the KV cache sharing and patterns, we describe the benchmark
set we build for \name. The study needs pairs of models that share the context
provided by the datasets. The following assumptions are also made when building the benchmark.
\begin{packeditemize}
      \item The pair of models should share the same foundational model.
        Specifically, the pair can either consist of the foundational model and
        a fine-tuned model based on it, or, two fine-tuned models based on the
        same foundational model.
  \item The dataset selected should be related to the task for which one of the
        models has been fine-tuned. This is important since in any
        context-sharing scenario, the receiver model is performing the
        specialized task.
  \item The receiver model fine-tuned on the task in the corresponding
        dataset should yield better quality than the sender
        model in the pair.
\end{packeditemize}

Using these assumptions, we formulate the benchmark as shown in
Table~\ref{tab:models}. We use 8 pairs of models across 6 datasets (including
HotpotQA~\cite{hotpotqa}, multifieldQA\_en~\cite{multifieldqa}, 
2wikimQA~\cite{2wikimqa}, multi\_news~\cite{fabbri2019multinewslargescalemultidocumentsummarization}, lcc~\cite{guo2023longcoderlongrangepretrainedlanguage},  and repobench-p~\cite{liu2023repobenchbenchmarkingrepositorylevelcode}). The quality metric used is taken
directly from the dataset.

We focus on the use case where the sender model 
generates the intermediate states for the context and the receiver model
reuses its intermediate states.
This is a challenging use case because the sender model has worse accuracy than the receiver model, so achieving high quality requires properly refreshing the KV cache. 





\begin{figure*}[ht!]
  \centering
  \includegraphics[width=0.85\linewidth]{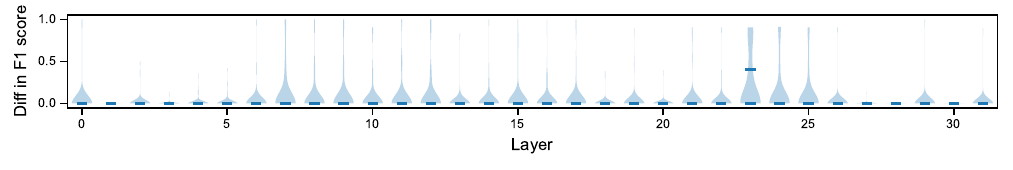}
  \vminten
  \tightcaption{
    Variation in F1 score per input within a single dataset (HotpotQA) for model pair Llama-3-8B-sft-lora-ultrachat reusing fingpt-llama-3-8B.
    We plot the 25 and 75 percentiles. Except layer 23, the 25 and 75 percentiles overlap, indicating a low variance of error sensitivity across all layers except 23.
  }
  \label{fig:layer_sens_violon}
\end{figure*}

\tightsubsection{Empirical insights of KV cache}
\label{subsec:empirical}
\mypara{Naive reusing is suboptimal}\label{subsubsec:locality}
The first observation is about naively reusing the sender model's KV cache on the receiver model. Specifically, we observe that:
\begin{insight}
\vspace{-5pt}
  Reusing the whole KV cache between models leads to a huge loss in accuracy.
\vspace{-5pt}

\end{insight}


A naïve way to reuse the intermediate state between models is to reuse the KV
cache {\em as is}. In this case, the receiver model receives the KV cache for the
whole input prompt from the sender model. It then uses this to generate the
output tokens in the decode phase, thereby skipping the prefill phase.

We show the impact of this on quality in Figure~\ref{fig:directreuse}. For
each pair of models and dataset, we show the F1 score (higher is better) of
{\em a)} the receiver model, {\em b)} the receiver model while reusing
the KV cache generated by the sender model, and {\em c)} the sender model
alone.

Although the quality of the receiver model with the sender model's KV
cache is still better than the sender model alone, we lose a lot of accuracy.
HotpotQA tends to lose more than 50\% of the accuracy points across most of the model pairs,
while the other datasets show varying amounts of changes across model pairs.

\mypara{Layer-wise sensitivity to KV cache reuse}
Our second observation is about whether KV cache reusing leads to the same
impact across all layers.

\begin{insight}
\vspace{-5pt}
  Only a small subset of layers are sensitive to KV cache reuse in terms of
  accuracy.
\vspace{-5pt}

\label{insight:second}
\end{insight}

Figure~\ref{fig:layerwise_pattern_across_models} shows the quality drop by
reusing part of KV cache from the sender  model. Specifically, each bar
represents the quality achieved by the receiver model reusing the KV cache
for that corresponding layer from the sender  model, with everything else
being recomputed.

For most of the model pairs, we find only a small subset of layers are
sensitive to the deviation in KV cache (\ie F1 score drops significantly), and
we refer to these layers as \emph{critical layers}, which are colored red. On
average across all pairs of models, we identify 11\% of layers to be critical. 

\mypara{Similarity of sensitivity across different inputs} Our third observation is about whether different inputs show similar patterns in
layer-wise sensitivity.

\begin{insight}
\label{insight:stable}
\vspace{-5pt}

  The variation in KV cache patterns across inputs is only notable for critical layers.
\vspace{-5pt}

\end{insight}

Figure~\ref{fig:layer_sens_violon} shows the violin plot of the normalized
change in F1 score per input in hotpotQA dataset, when
\texttt{llama-3-8b-sft-lora-ultrachat} reusing \texttt{fingpt-llama-3-8b}'s KV cache of each
layer only. Layer 23, which is also marked as the most critical for this model
pair in Figure~\ref{fig:layerwise_pattern_across_models} (\ie{} the largest F1
score change), shows a wider variation across different data points from the
dataset, with a lot of them observing F1 score change $> 50\%$. However,
for all the non-critical layers, the variance in the F1 score change is
insignificant, meaning that such non-critical layers do not change across
various inputs.

This phenomenon is also observed across other model pairs. Intuitively, this
can be because critical layers are essential for the reasoning capabilities~\cite{chen2020understanding} or the ability to accomplish specific downstream
tasks~\cite{chen2024biggerdeeperbetterprobing}. These reasoning capabilities must remain accurate to
interpret any input to the LLMs.


%% file: sec-design3.tex
\tightsection{\name Design}

Building on the insights in the previous section, we designed \name to enhance the context sharing between two LLMs. The central questions that \name targets are the following: {\em how do we determine the layers to re-compute to reduce latency, while keeping the quality loss minimal? }

\tightsubsection{Challenges with Selective KV Cache Reuse}

Insight~\ref{insight:second} suggests selectively reusing the KV cache while recomputing it for critical layers {\em might} preserve quality. However, selecting all critical layers scattered across different parts of the LLM is suboptimal for both efficiency and quality.

\mypara{The efficiency challenge} Recomputing critical layers that are non-contiguously placed is inefficient. 

During the prefill phase, the output of a layer where the KV cache is reused only contains information about the first generated token. In contrast, recomputing the KV cache needs to start from the E cache of this layer for the whole context. 
While it is possible to obtain the E cache for layer $l$ by performing a full prefill from the context starting from the first layer till layer $l$, this approach completely defeats the purpose of KV cache reuse.

\begin{figure}[t!]
  \centering
  \includegraphics[width=0.96\linewidth]{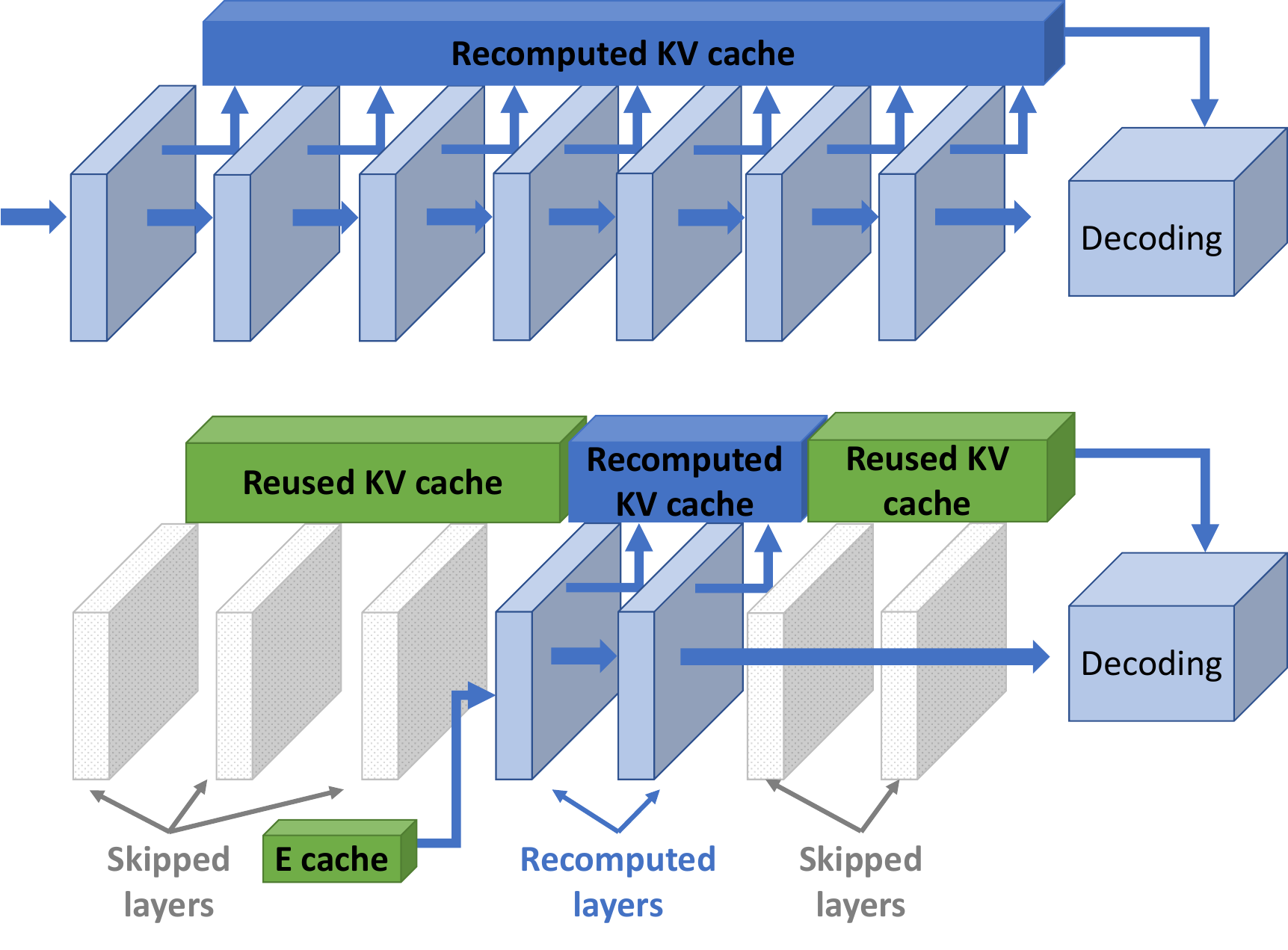}
  \vspace{-5pt}
  \tightcaption{
  \name chooses the critical layer groups (layers 4-5) to re-compute, and reuse KV cache for other layers. 
  }
  \label{fig:reusepattern}
\end{figure}

To address this, we use the E cache from the sender model to start the recomputing at the layer when transitioning from KV cache reuse to recompute.
We refer to this layer as a {\bf \em transition layer}. 
As depicted in Figure~\ref{fig:reusepattern}, for any transition layer (between reuse and recompute), the sender model must store and transmit the E cache to the receiver model.

The E cache is typically large, reaching up to \emph{twice} the size of the KV cache for the Mistral-7B or Llama-3-8B model families, and up to \emph{four} times larger for Llama-3.1-70B since the KV cache size is optimized by group-query attention~\cite{ainslie2023gqa}.
Thus, the overhead of storing the E cache in GPU memory and the delays caused by loading it from remote GPU nodes can be substantial, far exceeding the cost of storing and loading KV cache alone.

\mypara{The accuracy challenge} Furthermore, reusing the sender model's E cache at the transition layer also hurts the accuracy of the final output.
This is because the E cache loaded from the sender model (starting point of the recomputation) already differs from the receiver model.
Such difference eventually will introduce deviation from the point of recomputation and propagate over all later layers. It is crucial to minimize the error caused by such deviation.

If we pick all the critical layers, which are often not appearing in contiguous groups (Figure~\ref{fig:layerwise_pattern_across_models}), there will be multiple transition layers from reuse to recompute, introducing multiple deviations in E cache.  

\begin{figure}[t!]
  \hspace{25pt}\includegraphics[width=0.65\linewidth]{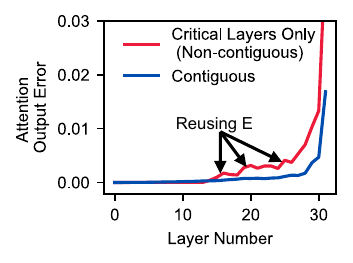}
  \vminten
  \tightcaption{
  More transition points lead to higher output error. 
  }
  \label{fig:cartoonfig}
\end{figure}

Figure~\ref{fig:cartoonfig} illustrates this. If we choose to recompute {\em only} critical layers (\ie layers 16--18, 20, and 25--27), we need to load E cache at layers 16, 20, and 25. 
However, whenever we load E cache, the error from E cache will be propagated to subsequent critical layers (\eg loading E cache at layer 16 populates errors to 16--18) and eventually to the output.
Thus, even if all the critical layers are recomputed, this will lead to a substantial output error.
In contrast, recomputing a contiguous group of layers from 16 to 27 avoids this problem by recomputing the KV cache of non-critical layers that are located between critical layers. 
As shown in Figure~\ref{fig:cartoonfig}, re-computing a contiguous group of layers has much lower output error than only re-computing the critical layers. 

\tightsubsection{Profiling for re-computation configuration}
\label{subsec:profile}
\begin{figure}[t!]
  \centering
  \hspace{-20pt}
  \includegraphics[width=0.58\linewidth]{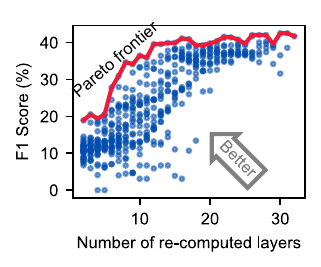}
  \vminfive
  \tightcaption{
  An example result obtained by offline profiling.  Each point represents the F1 score achieved when re-computing a specific group of contiguous layers. The Pareto frontier shows the maximum F1 score attainable for each number of re-computed layers. }
  \label{fig:heatmap}
\end{figure}

A key question still remains: how to determine the critical layer groups to be re-computed? 

Based on Insight~\ref{insight:stable}, the critical layers that are sensitive to KV cache differences vary little with the inputs.
This motivates our design to choose the critical layer group based on some example inputs, and then apply to other new inputs.   
Specifically, 
for each model pair, we use a ``training'' dataset to determine the critical layer group. 
We refer to the critical layer group as the \emph{re-computation configuration}. 
The goal is to understand the relationship between the critical layer groups to perform re-computation and its impact on generation quality.


Figure~\ref{fig:heatmap} shows an example profiling result for the \texttt{glue\_sst2} and \texttt{conllpp} model pair.
Each point in the scatter plot represents the F1 score achieved for a specific number of re-computed layers.
From these points, we obtain the Pareto frontier, which captures the highest F1 score achieved at a specific group of layers.
For instance, a configuration with 11 re-computed layers is a good example on the Pareto frontier, since the F1 score drop is within 5\%  of the original accuracy of the receiver model, while the number of re-computed layers is minimized.



\mypara{Profiling Overhead} The profiling overhead has a complexity of $O(l^2)$ where $l$ is the number of layers in the LLMs. 
For example, for \texttt{Llama-3-8B} with 32 layers, it takes three hours on an A100 GPU. This one-time cost is negligible since these models are deployed at a very large scale~\cite{patel2024splitwiseefficientgenerativellm, zhong2024distservedisaggregatingprefilldecoding}. Furthermore, this cost can be substantially reduced by grouping layers when profiling. Profiling at the granularity of 2-layer groups, for instance, reduces the profiling time by about 3x. 
In the evaluation we show soon in \S\ref{sec:eval}, we use the 2-layer group profiling granularity. 


\tightsubsection{\name runtime design}
\label{subsec:design}
\begin{figure}[t!]
  \centering
  \includegraphics[width=0.96\linewidth]{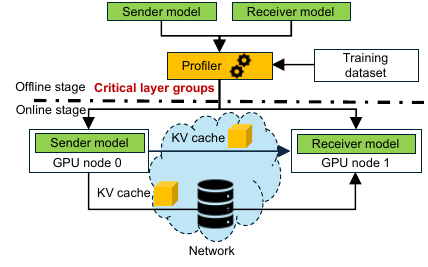}
  \vspace{-10pt}
  \tightcaption{
    Overall system design of \name, including the offline stage (top) that profiles the critical layers of each pair of sender and receiver models  (\S\ref{subsec:profile}); and the online stage (bottom) that uses the profile to dynamically pick the critical layers and KV-cache loading strategy for each job  (\S\ref{subsec:design}). 
  }
  \label{fig:recomp}
\end{figure}

As shown in Figure~\ref{fig:recomp}, \name consists of two stages. The offline stage (as detailed in \S\ref{subsec:profile}) uses a training dataset to profile the relationship between the value change of each layer on the generation quality. This step allows \name to dynamically choose the right re-computation configuration based on available resources. 
Next, we detail the {\em online} stage that uses the offline profile to dynamically decide the critical layers of the KV cache and executes the partial recomputation on these critical layers to preserve high generation quality. 
%


Specifically, in the online stage, \name dynamically decides which
point on the Pareto frontier should be used based on the latency SLO (details in Appendix\S\ref{appendix:slo}). 
We assume the E cache and KV cache at each transition point are precomputed and stored, but they can also be generated by the sender model in real-time and sent directly from the sender model.
Then, the receiver model selectively recomputes critical layer groups while reusing others, achieving a balance between computational efficiency and quality.

%



\begin{figure}[t!]
  \centering
  \includegraphics[width=\linewidth]{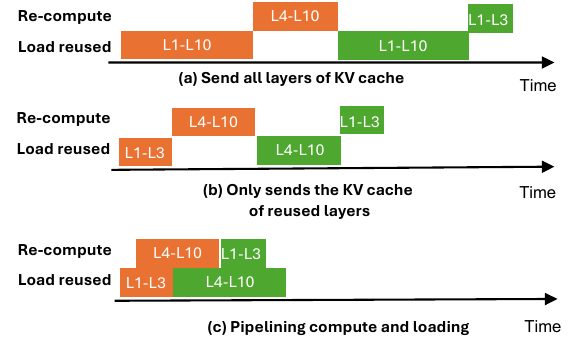}
  \vspace{-20pt}
  \tightcaption{
    Illustration of different KV cache transferring strategies. Each color represents the work (KV cache recomputation and loading reused KV caches) for the same query for one model. 
  }
  \label{fig:smart_loading}
  \vspace{-10pt}
\end{figure}

\mypara{Smart KV Cache Loading}
Since GPUs hosting different LLMs may reside in separate nodes, \name needs to fetch KV cache from remote nodes frequently. 
Without careful design, this transfer can significantly increase the end-to-end latency, particularly when fewer layers are recomputed (\ie more KV cache layers need to be transferred).

Consider an example with two receiver models $A$ and $B$.
Each model has ten layers. 
Model $A$ recomputes layers 4--10 and reuses KV cache for layers 1--3, while model $B$ recomputes layers 1--3 and reuses KV cache for layers 4--10.

Figure~\ref{fig:smart_loading} illustrates the timeline of three different loading and recomputation strategies. 
We assume that a request arrives at $A$ at time=0, followed by a request to $B$ after 2 time units, and that recomputing or transferring one layer (KV or E cache) takes 1 unit of time. 
The most naive approach is to load \emph{all layers} of the KV caches before recomputation begins in each job. 
For example, in model $A$ (orange), this means first loading the E cache for layer 4 and the entire KV cache, followed by 7 units of re-computation time, similarly for $B$, resulting in a total TTFT of 47 (Figure~\ref{fig:smart_loading}(a)).

A slight improvement is to load only the reusable layers’ KV cache for $A$ and $B$, right before recomputation. This reduces the total TTFT to 30, as shown in Figure~\ref{fig:smart_loading}(b).

However, both approaches miss the opportunity to overlap recomputation and the loading of reused KV caches. 
Recomputation can start immediately after receiving the E cache of the transition layer. The optimal solution, shown in Figure~\ref{fig:smart_loading}(c), pipelines loading and recomputation. For model $A$, the E cache for layer 4 is transmitted first, enabling recomputation of layers 4--10 to start while the KV cache for layers 1--3 transfers in parallel. 
Similarly, for $B$, the loading of E/KV cache can begin before $A$ finishes recomputation. This pipelined strategy reduces the total TTFT to 17—approximately a 2$\times$ improvement over the baseline in (b).

\tightsubsection{Implementation}
\label{subsec:impl}

We implement \name with about 3K lines of code in Python, based on PyTorch v2.0, CUDA 12.0, and LMCache 0.1.4~\cite{lmcache, cheng2024large}.
\name operates the LLM inference serving engines through these interfaces:
\begin{itemize}
  \item \texttt{store(Cache, context, LLM)}: We split the KV or E cache into layers, and store it in a key-value store in GPU memory.
  \item \texttt{fetch(context, LLM, layer\_id) -> Cache}: Depending on what was stored previous \texttt{store()} call, this loads layer's KV or E cache of the corresponding \texttt{LLM}.
  \item \texttt{partial\_prefill(recompute\_config, context)-> text}: it takes in the recomputation configuration and the context, including which layers to recompute during prefill, and then generates the output text.
\end{itemize}

We implement these three interfaces in vLLM~\cite{kwon2023efficientmemorymanagementlarge} and LMCache~\cite{lmcache}. For
\texttt{store\_kv}, after an LLM generates the KV cache for a piece of context,
we calculate the hash of the context text, and put it into the key-value store
if the context does not exist in the current store. Before we run the inference
for any LLM, we obtain the re-computation configuration from the offline
profiling (\S\ref{subsec:profile}), which includes the layer numbers for recompute and KV cache reuse.
During the online inference stage, we call the \texttt{partial\_prefill}
function, which calls \texttt{fetch\_kv} for the layers for KV cache reusing, and \texttt{fetch\_e} at the transition layers. 
Both \texttt{fetch\_kv} and \texttt{fetch\_e} are implemented with \texttt{torch.distributed}~\cite{PyTorch_Distributed_2024} to fetch KV or E cache from a remote GPU node.
All transmission will be placed on a CUDA Stream different from PyTorch's default computation stream \cite{pytorch_cuda_stream}, enabling us to overlap transmission of KV cache with recomputation and hiding the transmission delay.



%% file: sec-eval2.tex
\tightsection{Evaluation}
\label{sec:eval}
\begin{figure*}[t]
    \centering
    \includegraphics[width=0.72\linewidth]{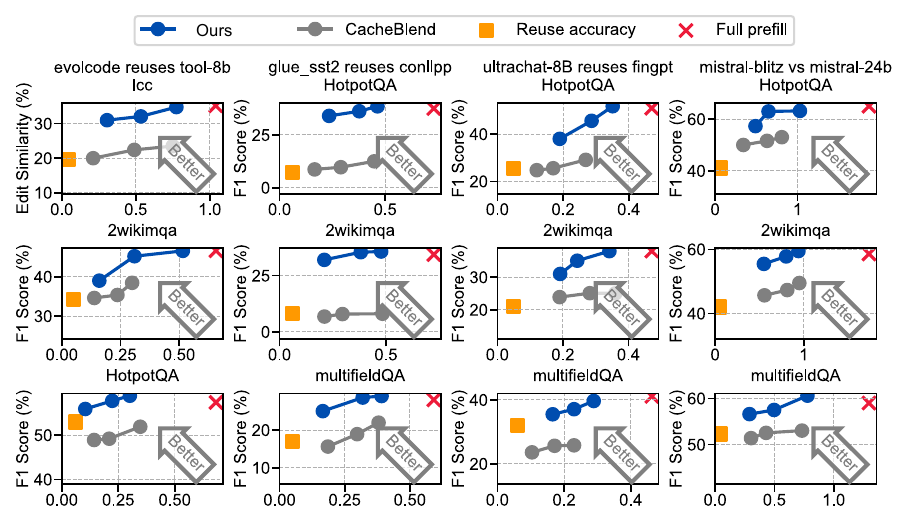}
    \vspace{-5pt}
    \includegraphics[width=0.72\linewidth]{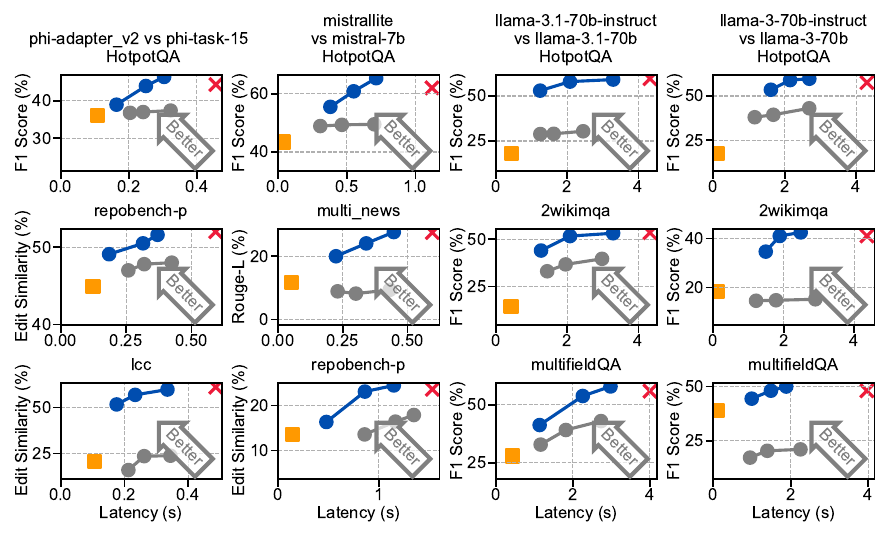}
    \tightcaption{Prefill delay and F1 score trade-off. \name greatly reduces prefill latency while maintaining generation quality.}
    \label{fig:eval}
\end{figure*}

\begin{figure*}[tbh]
  \centering
  \includegraphics[width=0.76\linewidth]{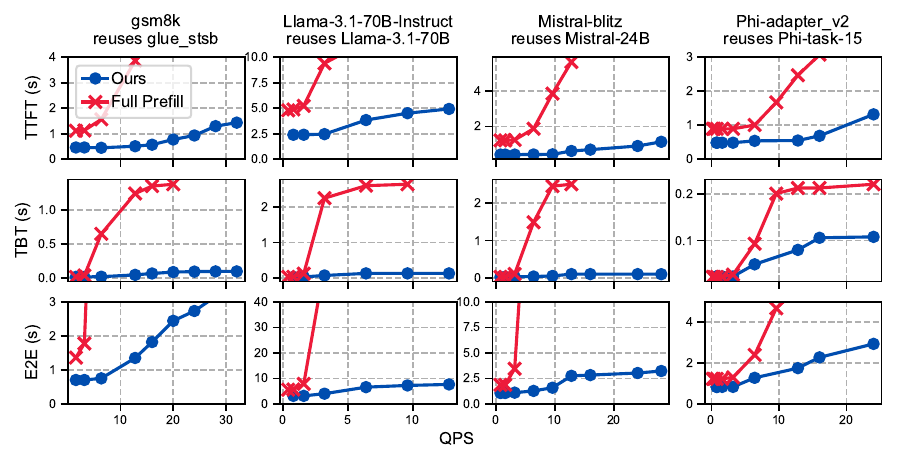}
  \vminten
  \tightcaption{
    The impact of arrival rate on Time-to-first-token (TTFT), Time-between-tokens (TBT), and end-to-end latency (E2E), when the \name's quality is same as full prefill. 
    Ran the models with eight replicas, with round-robin routing. 
    }
  \label{fig:qps-vs-time}
\end{figure*}

The key takeaways from the evaluation are:
\begin{itemize}
  \item Across three datasets and eight model pairs, \name can reduce the
        prefill latency by 1.7--3.1$\times$ without compromising accuracy, with an average prefill speedup of 2.1$\times$.
  \item In the online serving system, \name achieves up to 4$\times$ improvement in throughput. 
  \item \name's profiling of recomputing layers is robust across different datasets and model types.
\end{itemize}
\tightsubsection{Experiment Setup}

\mypara{Models} We evaluate \name on eight pairs of models (Table~\ref{tab:models}) of different sizes,
specifically the fine-tuned versions of Mistral-7B, Mistral-24B, Llama-3.1-8B, Llama-3-8B, Phi-3.5-mini-instruct, Llama-3-70B and
Llama-3.1-70B, selected with the criteria in \S\ref{subsec:bench}.
These models are fine-tuned on the base foundation model for chat-enhancing tasks, coding tasks, and long context reasoning \etal.
For Llama-3.1-70B and Llama-3-70B models, we use 4-bit quantized models with AWQ~\cite{lin2024awqactivationawareweightquantization} to fit on one A100 GPU. 

Note that the receiver models are not all directly fine-tuned from the sender model; they also include \emph{two fine-tuned variants} derived from the same foundation model.

\mypara{Hardware setting} We run the experiments on two A100 virtual machines connected with InfiniBand link on Microsoft Azure, namely \texttt{Standard\_ND96amsr\_A100\_v4}, which contains 8 $\times$ 80GB A100 GPUs on each virtual machine.

\mypara{Datasets} We evaluate \name on six datasets, with their context length statistics summarized in Table~\ref{tab:datasets}.
For each model pair, we report results on three of these datasets, following the criteria that the receiver model must achieve higher generation quality than the sender model on the chosen datasets.
The tasks drawn from the LongBench evaluation suite~\cite{longbench} evaluate LLM capabilities in multi-hop reasoning, summarization, and code understanding or completion.

\mypara{Train/test split} As discussed in \S\ref{subsec:profile}, \name profiles the critical layer groups that has quality drop within 5\% of the original quality with a ``training'' dataset offline.
Specifically, we use 50 contexts from HotpotQA dataset as the ``training'' dataset to obtain the critical layer groups with the profiling mechanism mentioned in \S\ref{subsec:profile}.
We apply the critical layer groups on all the other testing datasets in the benchmark.

\mypara{Quality metrics} We measure generation quality using the standard metric of each dataset, following prior work~\cite{longbench, liu2024cachegenkvcachecompression, yao2024cacheblendfastlargelanguage}. Specifically, we use F1 score for QA tasks (\texttt{hotpotQA, 2wikimQA, multifieldQA\_en}), which measures the probability that the generated answer matches the ground-truth answer for the question-answering task; Rouge-L score for summarization tasks (\texttt{multi\_news}), which measures the longest common subsequence between the generated summarization and the ground-truth answer; and finally the code similarity score for code completion tasks (\texttt{lcc, repobench-p}), which measures the edit distance between the generated completed code and the ground-truth code.

\mypara{System metrics} We use the system metrics listed in \S\ref{sec:transformer} to evaluate \name compared with the baselines, including TTFT, TBT, E2E.
In \S\ref{subsec:accuracy_latency_tradeoff} we also measure prefill latency, which includes the prefill computation time on GPU and the loading delay to fetch KV and E cache through InfiniBand bandwidth link across two GPU nodes.


\mypara{Baselines} We compare with the following baselines:
\begin{itemize}
  \item Full prefill: the receiver model prefills the text of the context with
        vLLM~\cite{kwon2023efficientmemorymanagementlarge}, representing the baseline of the highest computation overhead
        but the best quality we can get.
  \item Full KV cache reuse~\cite{gao2024costefficientlargelanguagemodel}: the receiver model directly reuses the KV cache
        from the sender model, and the receiver model runs decoding with the
        transferred KV cache.
\item CacheBlend~\cite{yao2024cacheblendfastlargelanguage}: we extend CacheBlend's algorithm to determine the important tokens to re-compute for cross-LLM KV cache sharing, based on the difference between the re-computed KV cache and sender model's original KV cache for the first layer.  
  \item Smaller models: In \S\ref{subsec:smaller_model}, we also compare
        \texttt{Llama-3.1-70B-Instruct}'s accuracy and latency trade-off with \name with
        \texttt{Llama-3.1-8B-Instruct}, which is fine-tuned with the same instruct-tuning
        dataset.
\end{itemize}

\tightsubsection{Lower Latency with Preserved Accuracy}
\label{subsec:accuracy_latency_tradeoff}
We first demonstrate \name's reduction in prefill delay and accuracy trade-off in Figure~\ref{fig:eval}. Across 8 pairs of models on three datasets, \name achieves 1.7--3.1$\times$ reduction in prefill
delay over the full prefill method, without compromising generation quality. On the other hand,  when compared with reusing all of
sender model's KV cache, \name successfully preserves the improved quality of the receiver model despite a slightly higher delay. 
Compared to CacheBlend, \name can achieve much better latency and quality trade-off. Specifically, \name has 5--33\% (average 16\%) higher quality than CacheBlend at a similar prefill latency. 

\mypara{Understanding \name's improvement} \name outperforms the baselines for various reasons. Compared to the full prefill baseline, \name achieves significantly lower prefill delay as only a small fraction of layers are prefilled. In contrast to full KV reuse, \name has a longer prefill latency because it does not perform prefill at all. However, it greatly reduces accuracy because it misses the opportunity to leverage layer-wise sensitivity in the KV cache difference.
\name is better than CacheBlend in terms of quality because \name re-computes the critical layer groups that are most sensitive to KV cache deviation between models, while CacheBlend fails to spot the critical layers since it selects the tokens to re-compute based on the first layer.



\tightsubsection{Throughput and Latency Improvement}\label{subsec:throughput-latency-improv}

To see the impact of \name on improving the throughput of an online LLM inference 
system, we emulated an online inference scenario by pairing the datasets with request arrival times following a Poisson distribution under different incoming rates to evaluate the performance of \name in more practical workloads. 

For the experiment, we deployed a Kubernetes cluster running vLLM Production Stack~\cite{vllm_production_stack_2025}, using \name’s customized Docker image. The cluster consisted of two nodes, each equipped with 8 A100 GPUs. We configured eight replicas for each model across these nodes. Model placement followed a simple strategy: for each model pair, we deployed four replicas of both the sender and receiver models on each node. 
Request routing was handled by vLLM Production Stack’s round-robin algorithm, which splits incoming requests evenly across all model replicas.

As demonstrated in Figure~\ref{fig:qps-vs-time}, we compare the TTFT, TBT, and E2E
impact under various request rates on the HotpotQA dataset. 
Due to the limit in space, we only show four pairs of
models to illustrate. For \name, we chose the configuration within 1\%  accuracy drop 
for these pairs of models. 

\mypara{TTFT}
Since the full-recompute baseline has much higher prefill latency than \name, the queuing delay affects (knee in the curve) its TTFT at a much lower QPS than what \name can support. 

\mypara{TBT and E2E}
Although we are only reducing the TTFT directly in \name, the second-degree effect through less interference and better scheduling brings down the TBT and E2E latency too, as shown in Figure~\ref{fig:qps-vs-time}.

\mypara{Throughput}
Assuming an SLO that avoids the effects of high queuing delays (knee of full prefill) on TTFT, TBT, and E2E latency, \name can support 2-4$\times$  higher throughput.

\begin{figure}[t]
  \centering
  \includegraphics[width=0.85\linewidth]{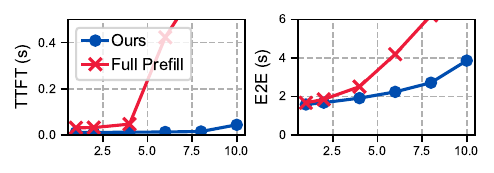}
  \vminten
  \tightcaption{
    Impact of the arrival rate on the Time-to-first-token (TTFT), and end-to-end latency (E2E) of code agentic workflow. 
  }
  \label{fig:agentic_workflow}
\end{figure}




\tightsubsection{Robustness across datasets}
\label{sec:robust}
As discussed in \S\ref{subsec:design}, \name profiles the KV cache reuse pattern 
using a single profiling run on a "training dataset" during the offline stage
and then generalizes the profile results to other datasets during the online stage.

\begin{figure}[t]
  \centering
  \includegraphics[width=0.9\linewidth]{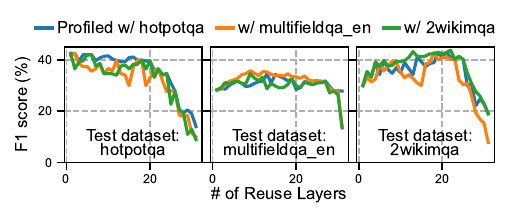}
  \vminten
  \tightcaption{
    Using the recompute layers profiled on training datasets works well on testing datasets.}
  \label{fig:robustness}
\end{figure}

Figure~\ref{fig:robustness} illustrates whether the profile obtained on one
dataset offline generalizes well to other datasets.
In each subfigure, we plot the Pareto frontier of the F1 score versus the
number of reused layers, obtained through profiling on the original testing
dataset vs two other datasets in our benchmark using \texttt{glue\_sst2} and \texttt{conllpp} model pair. 

The figure demonstrates that the Pareto frontier obtained using the profile
from the training dataset on the testing dataset closely resembles the frontier
obtained using the profile directly from the testing dataset. Across all the pertinent configurations, the maximum difference in the score is 4 points, with the average being 2 points. This result
further validates the sufficiency and robustness of our profiling strategy.

\tightsubsection{Case study of other tasks and other models}
\mypara{Math task} To demonstrate that the mechanisms of \name apply to other types of datasets, we apply \name on a model pair where the receiver model is fine-tuned on math reasoning, and test on a math reasoning task. 

\begin{figure}[ht]
  \hspace{-10pt}
  \includegraphics[width=\linewidth]{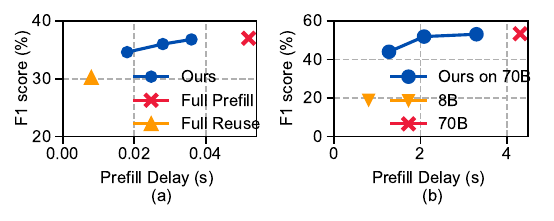}
  \vminten
  \tightcaption{(a) Prefill delay and accuracy trade-off for MAmmoTH2 (fine-tuned on math reasoning tasks). (b) \name applied on Llama-3.1-70B-Instruct has higher accuracy than Llama-3-8B-Instruct.
  }
  \label{fig:smaller}
\end{figure}

In Figure~\ref{fig:smaller}(a), we run GSM8K~\cite{yue2024mammoth2scalinginstructionsweb} dataset on MAmmoTH2~\cite{yue2024mammoth2scalinginstructionsweb}. 
Note that the Pareto frontier obtained follows a very similar pattern compared to the LongBench models and dataset, demonstrating the wide applicability of \name.

\mypara{Case study on coding agentic workflow} 
Next, we study how \name performs under a real agentic workflow by orchestrating a coding agent system using MetaGPT \cite{hong2024metagpt}, a state-of-the-art multi-agent framework. The system consists of two agents, a coder using evolcode model and a tester using tool-8b model. The coder is responsible for implementing Python functions according to the input prompt, and the tester is responsible for testing the coder's code and providing comments to the coder agent for the next round's modification. 

We send the problems from the HumanEval dataset at various rates. In Figure~\ref{fig:agentic_workflow}, we plot the TTFT and E2E impact under different QPS, similar to the setup in \S\ref{subsec:throughput-latency-improv}, where \name is plotted with the re-computation configuration that maintains the generation quality (pass@1 score of 52.5). 
\name significantly improves the TTFT by 2.7$\times$ and brings down the E2E delay to finish the problem as well.

\begin{figure}[ht]
\centering
\vspace{-10pt}
  \includegraphics[width=0.8\linewidth]{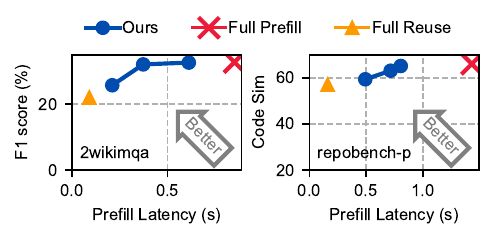}
  \tightcaption{
   Applying \name on Mixtral-8x7B-Instruct sharing Mixtral-8x7B's KV cache, which are MoE models based on Mistral model architecture. 
  }
  \vspace{-15pt}
  \label{fig:moe}
\end{figure}

\mypara{Mixture-of-Experts Model} We also evaluate \name on a Mixture-of-Experts (MoE) model.
As shown in Figure~\ref{fig:moe}, where the sender model is Mixtral-8x7B and the receiver is Mixtral-8x7B-Instruct, \name is able to achieve significant reductions in prefill latency as well.
While MoE models may activate different experts for different inputs, this selection occurs only at the linear layers after the attention modules—where the KV cache is used in.
As a result, \name’s KV cache sharing remains effective for MoE architectures.

\tightsubsection{Comparison against a smaller model}
\label{subsec:smaller_model}
Since \name trades off minimal accuracy impact for latency, we compare using \name on a larger model with a smaller model of the same architecture to show our superior performance in the quality and delay trade-off.

In Figure~\ref{fig:smaller}(b), we compare \name on \texttt{Llama-3.1-70B-Instruct}
and \texttt{Llama-3.1-8B-Instruct}, which is a smaller version of \texttt{Llama-3.1-70B-Instruct} and fine-tuned on the same dataset to enhance the base LLM's ability to follow instructions. As shown, \texttt{Llama-3.1-8B} achieves approximately a 4$\times$ reduction
in prefill delay but suffers a reduction in F1 score of about half 
compared to the original F1 score of \texttt{Llama-3.1-70B-Instruct}.

One significant drawback of using a smaller model to achieve speedup is the overhead of switching between small and large models. 
For example, when
additional resources become available, switching back to the larger model to
improve serving quality incurs the overhead of loading the larger model back
onto the GPU. 
In contrast, \name can easily adapt to the available compute resources by adjusting the
number of layers to be recomputed. This enables more possibilities for efficient scaling up or down on demand.

\begin{figure}[ht]
  \centering
  \includegraphics[width=0.9\linewidth]{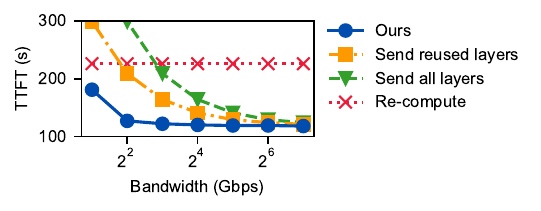}
  \vminten
  \tightcaption{
   Pipelining re-compute and loading the reused layers greatly reduces the total TTFT compared to the baselines of sequentially send and re-computes the KV cache. 
  }
  \label{fig:bandwidth}
\end{figure}

\tightsubsection{Impact of different network bandwidth}

Figure~\ref{fig:bandwidth} compares \name and the baselines where the re-computation and loading of KV cache are not pipelined.
We can see that \name outperforms baselines under almost all bandwidth situations.
Arguably, the absolute reduction in TTFT becomes smaller under high bandwidth, because the transmission delay becomes a smaller amount of the overall delay when the bandwidth is very high. 

%% file: sec-related2.tex
\vspace{-5pt}
\tightsection{Limitation}
\vspace{-5pt}
\mypara{KV cache sharing across different foundation models}
\name as-is does not support KV cache sharing across LLMs originating from different foundation models, whose resulting KV caches are of different sizes.
We leave this scenario to future work.

\mypara{Re-computation adaptation with network bandwidth}
In \S\ref{subsec:design}, we only consider adjusting the re-computation ratio based on system load.
Future work may extend the adaptation algorithm to consider changes in network bandwidth, for example, expanding the range of critical layer groups when network bandwidth is limited.

\mypara{Data drift in the re-computation configuration}
\name profiles the re-computation configuration for different models offline.
However, this approach may result in quality degradation if the real test data drifts significantly from the ``training'' data used for profiling.
Periodic re-profiling can be performed to update the re-computation configuration in line with new data.
We leave this enhancement to future work.

\vspace{-10pt}
\tightsection{Related Work}
\vspace{-10pt}
\mypara{Fine-tuning} Fine-tuning LLMs for specific tasks has gained importance, but it remains resource-intensive. Methods like parameter-efficient fine-tuning, including LoRA and LISA~\cite{pan2024lisalayerwiseimportancesampling,yao2024layerwiseimportancemattersmemory,dlora,sheng2024sloraservingthousandsconcurrent} reduce the memory and computation needed for fine-tuning. 

\mypara{Multi-agent systems} Multi-agent systems show promise in areas such as coding~\cite{holt2024l2maclargelanguagemodel, chen2024coderissueresolvingmultiagent, hong2024metagpt, microsoft_autogen, islam2024mapcoder, huang2024agentcodermultiagentbasedcodegeneration, qian2024chatdevcommunicativeagentssoftware}, gaming~\cite{zhang2024buildingcooperativeembodiedagents, gong2023mindagentemergentgaminginteraction, agashe2024llmcoordinationevaluatinganalyzingmultiagent, proagent, chen2024sagentsselforganizingagentsopenended, liu2024llmpoweredhierarchicallanguageagent, mosquera2024llmaugmentedautonomousagentscooperate}, and social simulations~\cite{10.1145/3586183.3606763, 51554}. Fine-tuned LLMs as agents improve outcomes in question answering~\cite{chen2023fireactlanguageagentfinetuning}, tool learning~\cite{shen2024smallllmsweaktool}, and personalization~\cite{li2024personalllmagentsinsights}. \name focuses on reducing communication delays in such systems.

\mypara{Faster LLM serving} One line of work speeds up LoRA model serving by hosting many LoRA models in memory at the same time. \name is faster than them due to the elimination of prefill computation. 
Other works improve LLM serving including better scheduling~\cite{agrawal2024tamingthroughputlatencytradeoffllm, sheng2024fairnessservinglargelanguage, miao2023spotserveservinggenerativelarge, kwon2023efficientmemorymanagementlarge,parrotserve, zhong2024distservedisaggregatingprefilldecoding, patel2024splitwiseefficientgenerativellm}, memory management for LoRA models~\cite{sheng2024sloraservingthousandsconcurrent, chen2023punicamultitenantloraserving}, and KV cache offloading~\cite{hu2024memservecontextcachingdisaggregated, gao2024costefficientlargelanguagemodel, jin2024ragcacheefficientknowledgecaching, infinigen}.
All of these works are orthogonal and complementary to \name.

Another closely related line of work also trades speed for quality but uses more compact model architectures~\cite{liu2024ghostnetv3exploringtrainingstrategies,sarah2024llamanasefficientneuralarchitecture,xia2020efficientsynthesiscompactdeep}. 
However, to smoothly adapt the amount of computation, they need to host multiple models of different sizes in GPU at the same time, which degrades the serving capacity in the system. 
\name does not suffer from it as it simply changes the number of recomputed layers.

\mypara{KV cache optimization} Lots of prior work has focused on optimizing KV caches for a single model. Some work focuses on compressing or offloading KV cache for reduced memory and transmission costs~\cite{zhang2023h2oheavyhitteroracleefficient, infinigen, oren2024transformersmultistaternns, xiao2024efficientstreaminglanguagemodels, xiao2024duoattentionefficientlongcontextllm}. Another line of research reduces the prefill delay when blending non-prefix KV caches from multiple contexts for the same model~\cite{yao2024cacheblendfastlargelanguage, gim2024promptcachemodularattention}. Since \name focuses on sharing KV cache across models, these works are orthogonal and can be used in conjunction.

%% file: sec-conclusion.tex
\vspace{-5pt}
\tightsection{Conclusion}
\vspace{-5pt}
In this work, we identified the core challenge of reducing repetitive computation in systems where multiple models work on a shared context.
We presented \name, a framework for
KV cache sharing in compound AI systems. We identified only a subset of layers that require recomputation to maintain quality and show the robustness of our solution for a diverse range of model pairs, model
types, and datasets.

%% file: sec-appendix1.tex
\section{Detail statistics of the datasets  }

\begin{table}[t!]
  \small
  \centering
  \begin{tabular}{ccccc}
    \toprule
    \textbf{Dataset}            & \textbf{Size} & \textbf{Med.} & \textbf{Std.} & \textbf{P95}  \\
    \hline
    hotpotQA~\cite{hotpotqa}          & 300           & 10933         & 5160          & 18650   \\
    2wikimQA~\cite{2wikimqa}          & 200           & 7466          & 3976          & 10705    \\
    multifieldQA\_en ~\cite{longbench} & 150           & 8084          & 3849          & 14680     \\
    multi\_news~\cite{fabbri2019multinewslargescalemultidocumentsummarization}  & 200 & 7624 & 5923 &   17547 \\
    lcc~\cite{guo2023longcoderlongrangepretrainedlanguage} & 200 & 12562  & 9220 & 26066 \\
    repobench-p~\cite{liu2023repobenchbenchmarkingrepositorylevelcode} & 200 &14285 & 8665& 30376 \\
    \bottomrule
  \end{tabular}
  \tightcaption{Size, context lengths, and evaluation metrics of datasets in the evaluation. }
  \label{tab:datasets}
\end{table}
Table~\ref{tab:datasets} shows the size of the datasets and the lengths of the contexts for these datasets.

\section{Dynamic Re-computation Configuration Adaptation}
\label{appendix:slo}

\begin{figure}[ht]
  \centering
  \hspace{-10pt}
  \includegraphics[width=\linewidth]{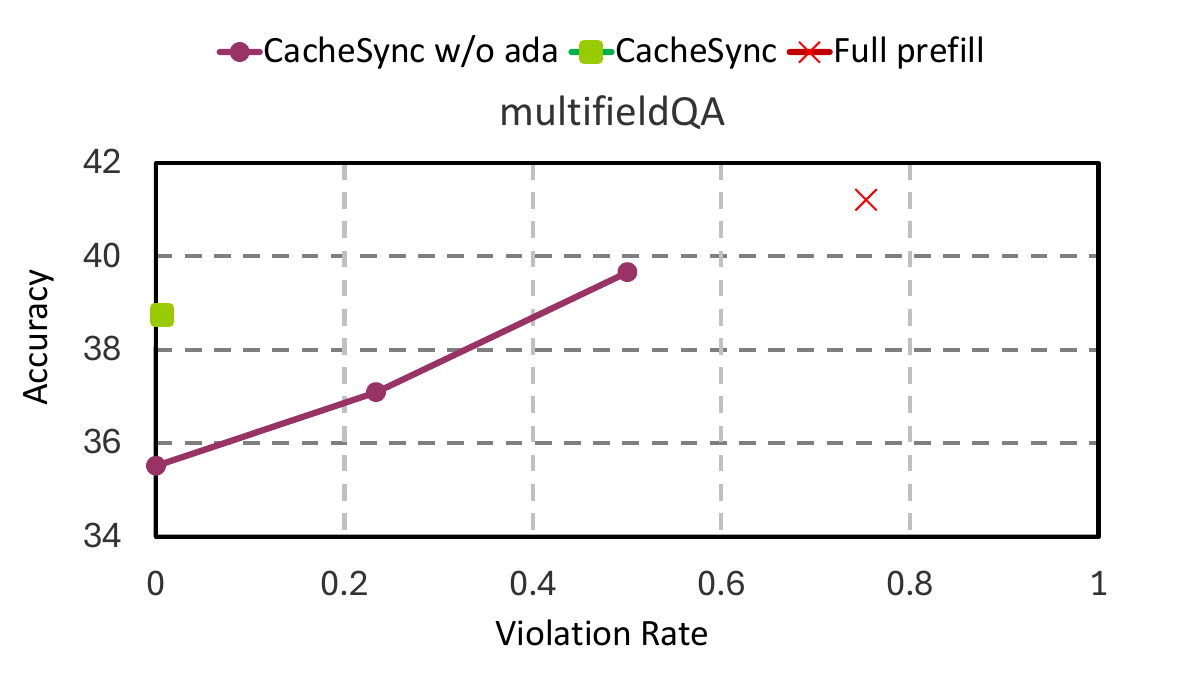}
  \tightcaption{\name reduces SLO violation rate over \name without adaptation and full prefill. Plotted with \texttt{ultrachat-8B} and \texttt{fingpt} model pair.
  }
  \label{fig:scheduling}
\end{figure}

\name initially checks the current workload intensity by monitoring the running and waiting requests in the vLLM engine~\cite{kwon2023efficientmemorymanagementlarge}. If there are requests that are waiting to be executed, it indicates that the current workload is high and triggers an increase in the reuse ratio. In this case, \name chooses a re-computation configuration that has the lowest number of re-computed layers above an accuracy target. When there are no queuing requests in the system, \name estimates prefill latency based on the number of tokens of each request. The parameters for estimation can be obtained through the profiling phase. \name then selects the highest recomputation ratio satisfying the latency SLO for each request to maximize accuracy. 